\documentclass[a4paper]{jpconf}
\usepackage{graphicx}
\usepackage[T1]{fontenc} \usepackage[latin1]{inputenc}
\usepackage{amssymb}
\usepackage{amsmath, amsthm,psfrag}

\newcommand{\va}{\scriptscriptstyle}

\newcommand{\R}{\mathbb{R}}

\newcommand{\Hk}{{\cal H}_{kin}}

\newcommand{\Q}{\mathrm{Q}}

\newcommand{\be}{\nopagebreak[3]\begin{equation}}
\newcommand{\ee}{\end{equation}}
\newcommand{\ba}{\nopagebreak[3]\begin{eqnarray}}
\newcommand{\ea}{\end{eqnarray}}

\newcommand{\bee}{\nopagebreak[3]\begin{equation*}}
\newcommand{\eee}{\end{equation*}}
\newcommand{\baa}{\nopagebreak[3]\begin{eqnarray*}}
\newcommand{\eaa}{\end{eqnarray*}}
\newcommand{\la}{\label}

\DeclareFontFamily{U}{rsfs}{}         
\DeclareFontShape{U}{rsfs}{m}{n}{<5> rsfs5 <6><7> rsfs7          %
  <8><9><10><10.95><12><14.4><17.28><20.74><24.88> rsfs10}{}     %
\DeclareMathAlphabet{\mathfs}{U}{rsfs}{m}{n}                     %
\begin{document}
\title{Non-commutative holonomies in $2+1$ LQG and Kauffman's brackets}

\author{K Noui$^1$, A Perez$^2$ and D Pranzetti$^3$}

\address{$^1$ Laboratoire de Math\'ematiques et Physique 
Th\'eorique, 37200 Tours, France} 

\address{$^2$ Centre de Physique Th\'eorique,
Campus de Luminy, 13288
Marseille, France}

 \address{$^3$ Max-Planck-Institut f\"ur Gravitationsphysik
AEI,
Am M\"uhlenberg 1, 14476 Golm, Germany}

\ead{pranzetti@aei.mpg.de}

\begin{abstract}
We investigate the canonical quantization of 2+1 gravity with $\Lambda>0$ in the canonical framework of LQG. A natural regularization of the constraints of 2+1 gravity can be defined in terms of the holonomies of $A_\pm=A \pm \sqrt\Lambda e$, where the SU(2) connection $A$ and the triad field $e$ are the conjugated variables of the theory. As a first step towards the quantization of these constraints we study the canonical quantization of the holonomy of the connection $A_{\lambda}=A+\lambda e$ acting on spin network links of the kinematical Hilbert space of LQG. We provide an explicit construction of the quantum holonomy operator, exhibiting a close relationship between the action of the quantum holonomy at a crossing and Kauffman's $q$-deformed crossing identity. The crucial difference is that the result is completely described in terms of standard SU(2) spin network states.\end{abstract}

\section{Introduction}

In his seminal work \cite{Witten1}, Witten established the link between the Jones Polynomial, Chern-Simons theory and
quantum gravity in 2+1 dimensions with $\Lambda\neq 0$. By defining a path integral quantization of the Chern-Simons theory with compact gauge Lie groups $SU(2)$, Witten provided a quantization
of Euclidean gravity with a positive $\Lambda$ and discovered a fascinating relation between knots invariants and the expectation values of Wilson
loops observables in Chern-Simons theory, leading to a new covariant definition of the Jones
polynomials.
After this result, Turaev and Viro defined a state-sum model \cite{Turaev:1992hq} showing the central role played by quantum groups in the construction of 3-manifolds invariants and knots polynomials. Turaev-Viro (TV) invariants can 
be viewed as a $q$-deformed version of Ponzano and Regge amplitudes. Moreover,
the asymptotic of the vertex amplitudes (the quantum $6j$-symbol)
has been shown to be related to the action of $2+1$ gravity with
non vanishing cosmological constant in the WKB approximation
\cite{Mizoguchi:1991hk}.
All this, strongly motivates the idea that it should be possible to recover (in
the context of LQG) the TV amplitudes as
the physical transition amplitudes of 2+1 gravity with $\Lambda\neq 0$. This has been so far explicitly shown only in
the simpler case for pure gravity with $\Lambda=0$
\cite{Noui:2004iy}. Following this simpler case, in LQG it is natural to interpret the TV
invariant as transition amplitudes between arbitrary pairs of spin-network states. For this to be the case, the TV amplitudes would have to be related to the
classical $SU(2)$
spin networks states of the canonical theory. In contrast, the TV amplitudes are constructed
from the combinatorics of $q$-deformed spin networks. This
would imply that the understanding of the relationship between the
TV invariants and quantum gravity requires the understanding of
the dynamical interplay between classical spin-network states and 
$q$-deformed amplitudes. We shall find here some indications of how
this relationship can arise. 

\section{Phase Space and Constraints}

Let us first briefly recall the canonical structure of (Riemannian)
gravity in 2+1 dimensions. The action of departure is
$S(A,e)=\int_{\mathcal{M}}\mathrm{tr}\left[e\wedge
F\left(A\right)\right]+\frac{\Lambda}{6}\mathrm{tr}\left[e\wedge
e\wedge e\right]$, where $\Lambda\ge 0$, $e$ is a cotriad
field, and $A$ is an $SU(2)$ connection.
Assuming that the space time manifold has topology $\mathcal{M}=\Sigma\times\R$, and, upon the standard 2+1 decomposition, the phase space of the theory is parametrized
by the pullback to $\Sigma$ of $\omega$ and $e$. In local coordinates
we can express them in terms of the 2-dimensional connection $A_{a}^{i}$
and the triad field $e_{a}^{i}$, where $a=1,2$ are space coordinate
indices and $i,j=1,2,3$ are $su(2)$ indices. The Poisson bracket
among these is given by $
\{ A_{a}^{i}(x), e_{b}^{j}(y)\} =\epsilon_{ab} \delta_{j}^{i}\delta^{(2)}(x, y)$. The phase
 space variables are subjected to the first class local constraints
\be d_Ae=0\ \ \ \ {\rm and } \ \ \ F(A)+\Lambda\ e \wedge e=0.
\label{constraints}\ee

The basic kinematical observables are given by the holonomy of the
connection and appropriately smeared functionals of the triad
field $e$. Quantization of these (unconstrained) observables leads
to an irreducible representation on a kinematical Hilbert space ($\Hk$) spanned by spin-network states. The holonomy acts simply by
multiplication while $e$ acts as the derivative operator.
Dynamics is defined by
imposing the quantum constraints on the
kinematical states. More precisely, the quantum
constraint-equations of 2+1 gravity with $\Lambda\neq 0$ can
be written as \be
\mathcal{G}\left[\alpha\right]\triangleright\Psi=\int_{\Sigma}\mathrm{Tr}[\alpha{\mathrm{d}_{A}e}]\triangleright\Psi=0
~~~ {\rm and}~~~
C_{\Lambda}\left[N\right]\triangleright\Psi=\int_{\Sigma}\mathrm{Tr}\left[N\left({
F(A)+\Lambda\, e\wedge e}\right)\right]\triangleright\Psi=0,
\label{Constraints}\ee
 for all $\alpha,N\in\mathcal{C}^{\infty}(\Sigma,su(2))$. Quantization of the previous expression for the constraints turns out to be difficult due to the presence of non-linear functional terms of the basic fields which requires the introduction of a regularization. 
A possible way to construct the physical Hilbert space of the theory is to follow the example of \cite{Noui:2004iy} for the case $\Lambda=0$. In fact, the main result of \cite{Noui:2004iy} is the definition of a
path integral representation of the theory from the canonical
picture reproducing the spin foam amplitudes
of the Ponzano-Regge model, and this sets the bases for the extension of
the analysis to the $\Lambda\neq 0$
case.
 Indeed, the key observation is that the curvature constraint in
(\ref{Constraints}) can be quantized by first introducing a
regulator consisting of a cellular decomposition $\Delta_{\Sigma}$
of $\Sigma$ into plaquettes $p$ and defining a new connection $A_{\pm}\equiv A\pm\sqrt{\Lambda} e$ so that we get
\be
C_{\Lambda}\left[N\right]=\lim_{\epsilon\rightarrow0}\sum_{p\in\Delta_{\Sigma}}\mathrm{Tr}\left[N_{p}\,
W_{p}\left(A_{\pm}\right)\right]-\mathcal{G}\left[\pm\sqrt{\Lambda}N\right],
\label{qcurvature}\ee
where $W_{p}(A_{\pm})\in SU(2)$ is
the Wilson loop computed in the fundamental representation. 
This provides a candidate background independent regularization of
the curvature constraint $C_{\Lambda}[N]$. The quantization of the previous classical
expression requires the quantization of the holonomy of $A_{\pm}$.
More generally, as a first step towards the quantization of (\ref{qcurvature}), we are now going to present the quantization of the holonomy $h_{\lambda}$ of the general
connection $A_{\lambda}\equiv A+\lambda e$ for $\lambda\in \R$, surveying the main steps of the more detailed analysis presented in \cite{QH}.

\section{Quantization of $h\left(A_{\lambda}\right)$}

We are now going to explore the quantization of the kinematical observable
$h_{\eta}\left[A_{\lambda}\right]={P}\
\mathrm{e}^{-\int_{\eta}A+\lambda e}$
associated with a path $\eta\in \Sigma$, as
operators on the $\Hk$ of 2+1 LQG.
In order to do this, let us first recall that in LQG there is a well-defined quantization of the $e$-field based
on the smearing of $e$ along one dimensional paths. More precisely,
given a path $\eta^a(t)\in \Sigma$ one considers the (smeared) connection conjugate momentum $
E(\eta)\equiv\int E^{ai}
\tau_i n_a dt, $ where $n_a\equiv
\epsilon_{ab}\frac{d\eta^a}{dt}$ is the normal to the path. The quantum operator associated to $E(\eta)$ acts
non trivially only on holonomies $h_{\gamma}$ along a path $\gamma\in
\Sigma$ that are transversal to $\eta$. It suffices to give its
action on transversal holonomies that either end or start on $\eta$.
The result is: 
\ba \label{flux}\hat E(\eta)\triangleright h_{\gamma}=\frac{1}{2}\hbar
\left\{\begin{array}{ccc} \!\! o(p)\tau^i\otimes \tau_i h_{\gamma}\ \ \mbox{if $\gamma$
ends at $\eta$}\\ o(p) h_{\gamma}\tau^i\otimes \tau_i\ \ \mbox{if $\gamma$ starts at
$\eta$}\end{array}\right., 
\ea
 where $o(p)$ is the orientation of
the intersection $p\in \Sigma$.
Due to the tensorial form of the Poisson bracket between the connection $A$ and the triad field $e$
(inherited by the commutator in the quantum theory) the action of
$h_\eta$ on the vacuum simply creates a Wilson line excitation,
i.e. it acts simply by multiplication by the holonomy of $A$ along the
path, namely $h_{\eta}[A_{\lambda}]|0\rangle=h_{\eta}[A]|0\rangle$. 

Therefore, the simplest non-trivial example is the action of the holonomy of $A_{\lambda}$ on a
transversal Wilson loop in the fundamental representation. We
define the quantization of $h_\eta$ by quantizing each term in its series expansion in $\lambda$. Terms
of order $n$  have $n$ powers of the $e$ operators. The
quantization of these products becomes potentially ill-defined due
to factor ordering ambiguities. 
More precisely, let $\gamma,\eta:(0,1)\rightarrow\Sigma$ be two curves
that cross each other \emph{transversally}. Let us now study
the action of $\eta$ on $\gamma$. Developing in powers of $\lambda$ the action
$h_{\eta}\left(A_{\lambda}\right)
\triangleright h_{\gamma}\left(A_{\lambda}\right)\left|0\right\rangle$, the generic coefficient of order $p$ is:
\ba \nonumber && \lambda^p\sum_{n\geq p}\,\sum_{m\geq
p}\,\left(-1\right)^{m+n}\,\sum_{1\leq k_{1}<\cdots<k_{p}\leq
n}\,\int_{0}^{1}\mathrm{d}t_{1}\cdots\int_{0}^{t_{n-1}}\mathrm{d}t_{n}\,\int_{0}^{1}\mathrm{d}s_{1}\cdots\int_{0}^{s_{m-1}}\mathrm{d}s_{m}\\
\nonumber &&
\left[A\left(\eta\left(t_{1}\right)\right)\cdots E(\eta(t_{k_1}))\cdots
E(\eta(t_{k_p}))\cdots
A\left(\eta\left(t_{n}\right)\right)\right]\,\triangleright
A\left(\gamma\left(s_{1}\right)\right)\cdots
A\left(\gamma\left(s_{m}\right)\right)\,.\label{action}\ea 
Ordering ambiguities now arise when computing the action of the derivation operators on the connection
along $\gamma$. In fact, it is easy to see that only the terms containing $p$ consecutive graspings $E$s acting on $p$ consecutive $A$s 
survive and this is where factor ordering ambiguities appear due to the non-commutativity of the grasping operators. We will see in a moment that relationship with Kauffman's bracket is found if one uses 
the so-called Duflo map. To lighten the presentation, let us now introduce a graphical notation for the quantum holonomy action 
\be\label{zz}
\begin{array}{c}\psfrag{a}{$ $}\psfrag{b}{$ $}
\includegraphics[width=1cm]{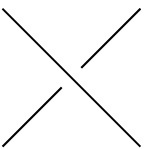}\end{array}=\begin{array}{c}\psfrag{a}{$Q_S$}
\includegraphics[width=1cm]{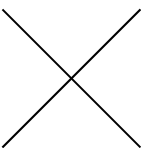}\end{array}+ z \begin{array}{c}
\includegraphics[width=1cm]{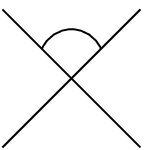}\end{array}+\frac{z^2}{2}\begin{array}{c}\psfrag{a}{\!$ \va Q_S$}
\includegraphics[width=1cm]{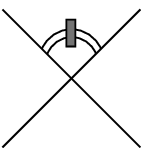}\end{array}+\frac{z^3}{3!}\begin{array}{c}\psfrag{a}{\!$\va Q_S$}
\includegraphics[width=1cm]{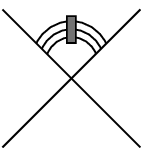}\end{array}+\cdots,
\ee where $z=-io\hbar \lambda$, and  the boxes, at this stage, denote a generic ordering prescription. At this point, inspired by \cite{Sahlmann}, we introduce the Duflo isomorphism as a quantization map and explicitly work out the prescription leading, once the action (\ref{flux}) of the flux operator in LQG is also taken into account, to a crossing evaluation in accordance with the $q$-deformed binor identity. 

The Duflo map \cite{Duflo} ($Q_D$) is a generalization of the universal {\it quantization map} proposed by Harish-Chandra for semi-simple Lie algebras. 
The latter provides a prescription to quantize polynomials of commuting variables (the classical triad fields $e$) which after quantization acquire Lie algebra commutation relations (the flux operators $\hat E$ in (\ref{flux})). We write explicitly only the expression of the map $Q_D$ for a pair of $E$ fields since, once taken into account the LQG quantization of fluxes described above, this is the only one we are going to use in the following, namely
\be\la{tautau}
\Q_D[E_jE_k]=\frac{1}{2}(\tau_j\tau_k+\tau_k\tau_j)+\frac{1}{6}\delta_{jk}.
\ee
In order to get the general form of the series (\ref{zz}) in the
case where we use the LQG quantization of the flux operators, it suffices to write the first few terms. In
the first order term, $E$ acts as a LIV on the portion of the
holonomy which has the crossing as its source and as a RIV on the
other one. The full result is just the same as in (\ref{zz}). In the second order diagram we have the action of two flux operators at the same point and therefore ordering ambiguities arise. In order to deal with them, we now use the prescription (\ref{tautau}) induced by the Duflo map and, diagrammatically, for the second order term we have
 $
\begin{array}{c}\psfrag{a}{\!$\va Q_{\va D}$}
\includegraphics[width=0.6cm]{x2.eps}\end{array}=\frac{1}{2}\begin{array}{c}
\includegraphics[width=0.6cm]{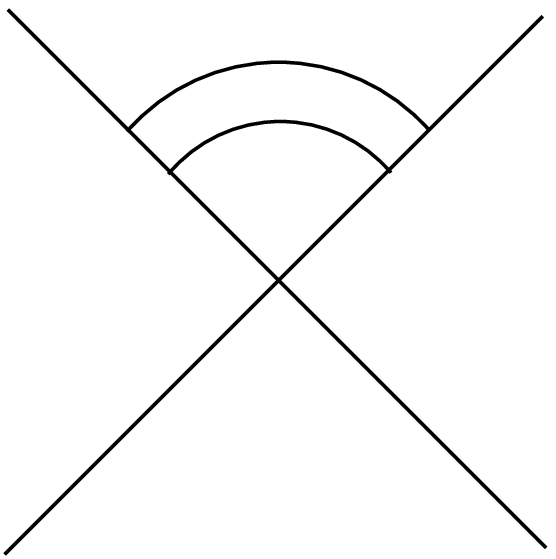}\end{array}+\frac{1}{2}\begin{array}{c}
\includegraphics[width=0.6cm]{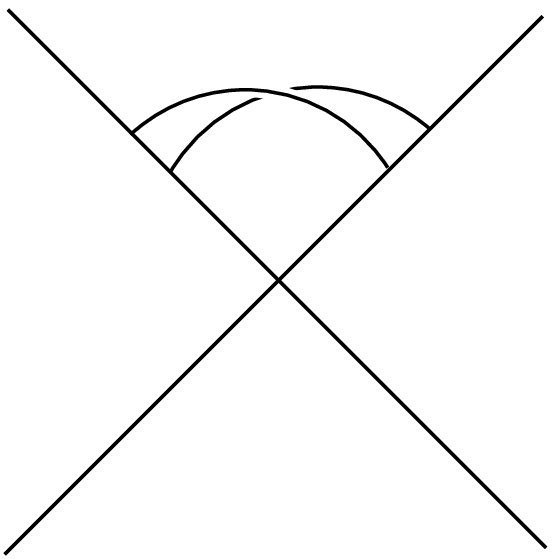}\end{array}+\frac{1}{6}\begin{array}{c}
\includegraphics[width=0.6cm]{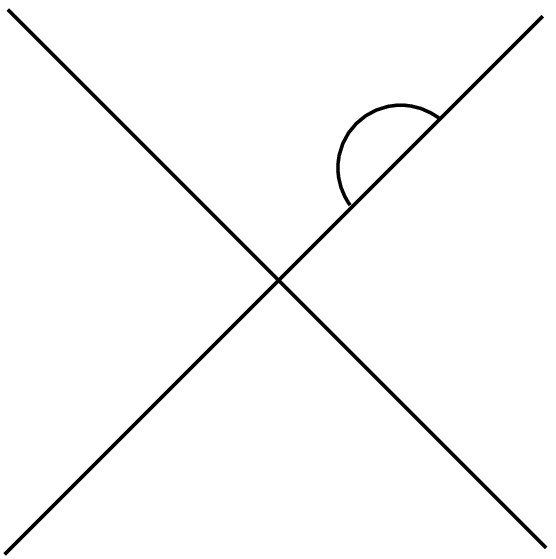}\end{array}=
\frac{1}{16}\begin{array}{c}
\includegraphics[width=0.6cm]{x.eps}\end{array}.
$ 
Therefore,
the second order diagram is proportional to the order zero
diagram. The third order
term is consequently proportional to the first order one and so
on. We get in this way the general expression for arbitrary order and perturbative expansion (\ref{zz}) can be exactly summed, yielding
\be
h_{\eta}\left(A_{\lambda}\right)\triangleright
h_{\gamma}\left(A_{\lambda}\right)\,\left|0\right\rangle=
\begin{array}{c}\psfrag{a}{}\psfrag{b}{}
\includegraphics[width=0.8cm]{x-quantum.eps}\end{array}=A  \begin{array}{c}
\includegraphics[width=0.8cm]{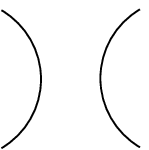}\end{array}+A^{-1}\begin{array}{c}
\includegraphics[width=0.8cm]{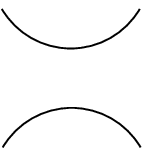}\end{array}
\label{kau},\ee
where $A=e^{\frac{i o\hbar \lambda}{4} }$. 
Therefore, the series expansion in powers of $\lambda$ converges and leads to a simple
expression for the crossing reproducing Kauffman's $q$-deformed binor identity
for $q=\exp{i\lambda/2}$.

\section{Discussion}\label{discu}

We have shown that the ordering ambiguities arising in the quantization of
the holonomy of $A_\lambda$ in the fundamental representation
can be sorted out by means of a simple quantization based on the Duflo map, leading to the Kauffman-like algebraic structure 
for the action of the quantum holonomy defining a crossing. 
The recovering of Kauffman's bracket related to the $q$-deformed crossing identity
is a remarkable result since it was obtained starting from the standard $SU(2)$ $\Hk$ of LQG.
This result is promising in the road to finding
a relationship between TV amplitudes and physical amplitudes in canonical LQG. In fact,
recall that the value of $d_q$ together with the
deformed binor identity are the two ingredients for the
combinatorial definition of the TV invariant. Despite the fact that, at the present kinematical level, loops still evaluate according to the classical $SU(2)$ recoupling theory, an intriguing indication  that the implementation of dynamics could lead to the emergence of the quantum dimension for loops evaluation emerges from the study
of the algebra of the operator (\ref{qcurvature}), taken as a proposal for the regularized version of the curvature constraint in (\ref{Constraints}). More precisely, the classical constraint algebra dictates that this should be  proportional to the Gauss constraint. If one performs this analysis, it is immediate to see that, besides the `mildly' anomalous contributions found in \cite{Anomaly} (see \cite{Ansatz} for a possible way around this difficulty), there are stronger ones which do not annihilate  gauge invariant states. Surprisingly, the latter anomalous terms happen to be proportional to $(A^2+A^{-2}+\begin{array}{c}  \includegraphics[width=0.3cm,angle=360]{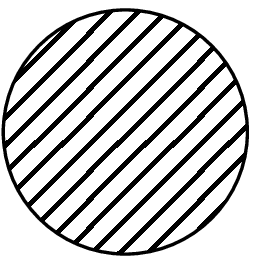}\end{array})$, where $\begin{array}{c}  \includegraphics[width=0.3cm,angle=360]{0loop.eps}\end{array}$ represents the loop with no area in the fundamental representation $j=1/2$. Thus the condition that an infinitesimal loop evaluates to the quantum dimension emerges from the constraint algebra.
 All this indicates that, even when we do not introduce a quantum group at any stage, and no
dynamical constraint has been imposed yet, amplitudes such as the
value of the quantum dimension 
and $q=A^2=e^{\frac{i\hbar\lambda}{4}}$ naturally appear from our
treatment. 
This indicates
that perhaps a strict correspondence between LQG and the
TV invariant can be established if one appropriately
quantizes and imposes the curvature
constraint (\ref{Constraints}). Quantum holonomies defined here might be the right tool along this direction.


\end{document}